# Multiplexed plasmonic nanoantennas for high throughput single molecule nanoscale dynamics in living cells

Pamina M Winkler[1#*], Thomas S van Zanten[2*], Satyajit Mayor[2&], and Maria F Garcia-Parajo[1, 3&]

[1]ICFO-Institut de Ciencies Fotoniques, The Barcelona Institute of Science and Technology, 08860 Barcelona, Spain

[2]National Centre for Biological Sciences, Tata Institute of Fundamental Research, Bellary Road, Bangalore, India

[3]ICREA, Pg. Lluis Companys 23, 08010 Barcelona, Spain

*Contributed equally

# Current address: Institute of Materials, EPFL, Lausanne, Switzerland

[&] Equally corresponding authors.

Corresponding authors e-mail addresses:

Maria.garcia-parajo@icfo.eu

mayor@ncbs.res.in

**Single molecule detection has revolutionised the fields of chemistry and biology by offering powerful ways to study individual molecules under different scenarios[1–3]. Nanophotonic structures[4,5], including plasmonic antennas[6–9], significantly overcome the concentration limit at which single molecule events can be observed, enabling their detection at concentrations that are relevant to biological and chemical processes[10–13]. Although antennas can be fabricated in large arrays[9,14–16], probing dynamic events requires high temporal resolution, which is best achieved by serial antenna interrogation[8,9,11,17]. Unfortunately, this precludes the simultaneous recording from multiple antennas at different sample locations, and is time consuming, resulting in poor statistics and low-throughput data acquisition, abating the true potential of arrays. Here we exploit arrays of antenna-in-box nanostructures in combination with sCMOS readout to interrogate nanoscale volumes from 225 antennas simultaneously. Recording at 1 kHz allowed multiplexed dynamic measurements from 50 nanoantennas simultaneously with a temporal resolution dictated by the camera framerate and the photons emitted per molecule during a single passage. We demonstrate the capability for high-throughput arrayed detection of single molecule dynamics at the nanoscale in the membrane of living cells, and determine the axial location of membrane molecular components with 1 nm accuracy and temporal resolution below 1 ms.**

The antenna design is based on the "antenna-in-box" concept[8], which optimally enhances single molecule fluorescence at the gap region and minimizes background from direct excitation of molecules away from the antenna gap. Arrays of several thousands of gold antennas-in-boxes with reproducible gap sizes in the range of 10-35 nm were fabricated on standard glass coverslips using a multistep process based on electron-beam lithography and template stripping to render them planar[14]. The planarization step guarantees that the region of maximum field enhancement is located upright and close to the sample[14,17,18]. Antenna arrays were placed on an inverted microscope and excited in an epi-configuration at the nanoantenna resonance wavelength of 640 nm in the presence of freely diffusing fluorescent beads (Fig. 1A). Fluorescence signals from 225 different nanoantennas were recorded with a sCMOS camera operating at a 10 Hz framerate and 5 ms integration time. Each antenna appeared as a separate bright diffraction-limited spot in the field-of-view (FOV), occupying several pixels of the camera (Supplementary Fig. 1A). Pixel-specific parameters were extracted as described[19] and used to convert the signal into photo-electrons. The total fluorescence per antenna at each timeframe was extracted by integrating the signal of all the pixels within a circle of radius $r$ around the centre of the nanoantenna. With $r$ = 4 pixels we obtained $3 \cdot \sigma \sim$ 99.7% spatial coverage of the signal emitted by a nanoantenna (Gaussian fitting results in $\sigma$=1.47 pixels, Supplementary Fig. 1A, Supplementary Note 1).

To assess the performance of the multiplexing scheme, we first addressed two key nanoantenna properties: polarisation-dependent resonance and fluorescence enhancement as a function of the antenna gap size. For this, we simultaneously recorded the fluorescence emission of 45 nm beads diffusing in a viscous environment from multiple nanoantennas. The increased viscosity slows beads diffusion to $\sim$ 1.4±0.5 $\mu m^2$/s (Supplementary Fig. 1B) thereby increasing the number of detected photons while diffusing through individual nanoantennas. High-intensity bursts were detected only when the antennas were excited with polarisation parallel to the gold dimer, and not when orthogonal polarisation was used (Fig. 1B), confirming their resonance behaviour. At 24 nM bead concentration, approximately 12 high-intensity bursts per trace were detected (Supplementary Fig. 1C) and the signal was largely dominated by background. Normalising to the median fluorescence intensity rescales the photon distribution histogram and its standard deviation associated with Poisson photon statistics to unity. This allowed a direct comparison between different antennas and showed the increased probability of intense fluorescence bursts as function of gap size, demonstrating our ability to measure nanoantenna enhancement using sCMOS detection (Fig. 1C). To evaluate the performance of each of the 225 nanoantennas we generated a "peak-performance map" (Fig. 1D). Every pixel in the map represents an individual antenna, pseudo-coloured according to its largest intensity event during the measurement. Fig. 1D confirms the clear trend of increasing intensity (brighter pixels) upon decreasing gap size.

Because the peak performance value of each antenna corresponds to a single bead diffusion event, the associated intensity can be converted into a relative enhancement with respect to the median performance of the apertures (300×140 $nm^2$ boxes lacking the antennas). Collecting more than 3000 events from 225 antennas per gap size, allowed us to extract the top 1% best-performing nanoantennas. We find a maximum relative enhancement $\sim$ 20-fold for the smallest gap antennas (Fig. 1E). Although modest as compared to other reports[6,8,9,14,20] the enhancement here is dominated by the increased near-field excitation with little or no contribution from further increase on the beads quantum yield ($\Phi$=0.9). Moreover, the enhancement is calculated using a fixed camera integration time and is therefore underestimated since a diffusing bead spends a shorter time in a smaller gap and will thus emit proportionally fewer photons. The photon budget underestimation due to shortening of diffusion times scales by 1.37-, 1.85-, 2.63- and 4.35-fold for gap sizes of 30 nm, 25 nm, 20 nm and 10 nm, respectively. The antenna gap size-dependence on fluorescence enhancement has been already demonstrated from hundreds of repetitive rounds of serial sampling of arrayed gap antennas[14]. Here, we validate this trend by one single multiplexed measurement on 225 nanoantennas per gap size. With only five independent measurements (corresponding to five different gap sizes), we recorded statistics on 1125 different antennas, corresponding to a 225-fold increase in data-acquisition throughput.

Our temporal resolution is dictated by the camera integration time and electronics readout rate. Reducing the FOV to about 50×950 pixels, corresponding to 6.1×115.9 μm² in the image plane, allowed recording of 50 nanoantennas simultaneously at a framerate slightly faster than 1 kHz (Fig. 2A). Collecting $2.4 \cdot 10^4$ frames resulted in a ~24s fluorescence trace, $F(i,t)$, for each nanoantenna $i$. Traces were analysed using fluorescence correlation spectroscopy (FCS) to extract the diffusion of individual beads at the antenna gap[21]. FCS computes the temporal autocorrelation function (ACF) of the fluorescence signal $G(i,\tau) = \langle F(i,t) \cdot F(i,t+\tau) \rangle / \langle F(i,t) \rangle^2$, where $\tau$ is the delay (lag) time and $\langle \rangle$ indicates time averaging[21]. In the absence of uncorrelated signal, the amplitude of the correlation function $G_{00}$ at zero lag time scales with the inverse of the average number of molecules $N$ in the detection volume. In the current scheme, however, the detected signal is the sum of correlated signal emanating from the nanoantenna, uncorrelated intensity from reflected red-shifted photons from the thin gold film[22] and pixel-associated readout noise. The bead brightness, $Q = \langle F \rangle / N$, as function of pixel radius $r$, increases upon increasing $r$ to a maximum at $r$ = 3-pixels, above which the signal decreases as the fraction of correlated photons from the nanoantenna reduces with respect to the uncorrelated background signal (Fig. 2B). Hence the optimum interrogation area on the sCMOS camera corresponds to $r$ = 3-pixels. This improved signal-to-noise is both crucially dependent on the nanoantenna being in resonance (Supplementary Fig. 2B) and the correct determination of its centre-of-mass, as the signal from the gold film is only uncorrelated background (Fig. 2C). As such, the antenna enhancement drastically increases the photon budget and should allow for multiplexed FCS measurements at the nanoscale with camera recordings at framerates above 1 kHz.

As a proof-of-principle, we recorded the diffusion of fluorescence beads over sets of 50 antennas simultaneously excited, and generated ACF-carpets from hundreds of individual ACFs (Fig. 2D). The ACF-carpet qualitatively displays faster decaying ACF curves and larger correlation amplitude values as the gap sizes decrease (Fig. 2D), both being strong signatures of the increased volume confinement[8,14,23]. Data quantification indeed rendered increased bead brightness (Fig. 2E) and shorter diffusion times (Fig. 2F) as the gap sizes reduce[8,14] reflecting the increased enhancement and spatial confinement brought about by the antennas. Notably, these results obtained via FCS dynamics are fully consistent to those obtained by burst analysis (Fig. 1E), endorsing the validity of our approach for multiplexed nanoscale-FCS studies with high temporal resolution. Importantly, nanoantenna properties such as resonance-dependent excitation polarisation, localised fluorescence enhancement, deterministic volume reduction and efficient background screening outside the nanoantennas are fully preserved.

To evaluate our method in biological relevant scenarios, we then measured the dynamics of individual molecules in the plasma membrane of living cells. Cell adhesion on the nanoantenna array was favoured by briefly exposing the array to UV-plasma treatment. Next, AGS cells overexpressing the GPI-anchored folate receptor (FR-GPI) spread on the array in about two hours after which the receptor was labelled with Abberior635-conjugated Fab-fragments. The cell contours were determined using a brightfield lamp from above the sample and the antennas over which the cells were spread lit up when excited (Fig. 3A). This indicated that there is little (if any) non-specific Fab-fragments sticking or diffusing through antennas away from the cells such that the collected signals are specifically associated with proteins diffusing in the cell membrane above the antennas. The signals also indicate that the receptors are in close proximity to the nanoantennas excitation field, allowing simultaneous recording of individual diffusing proteins in living cells. The multiplexed ACF-carpet generated from 51 ACF traces was markedly distinct from that generated on diffusing beads (Fig. 3B). Firstly, the ACF curves were shifted towards longer timescales, consistent with slower protein diffusion on the membrane, as compared to beads diffusing in solution. Secondly, the curves revealed a large dynamic heterogeneity within similar gap sizes. Indeed, ACFs curves of FR-GPIs on antennas of similar gap size exhibited markedly different correlation amplitudes and diffusion times (Fig. 3C). This heterogeneity has been documented before and is likely due to GPI-APs undergoing nanoclustering and temporal arrests, characteristics that become most prominent when observing diffusion at smaller spatial scales[24–26]. Thirdly, and surprisingly, there was no clear trend associated with the antenna gap sizes (Fig. 3B). We thus generated a scatter plot of the relative brightness and diffusion times as extracted from individual ACF curves recorded on antennas regardless of their gap size, and compared them to the values obtained on apertures (Fig. 3D). In contrast to the apparent averaged outcome when measuring FR-GPI mobility in apertures, the measurements on nanoantennas were highly dispersed. We observed a similar trend when measuring a GPI-AP diffusion on a different cell type (Supplementary Fig. 3A-B). The variation in both the timescales and brightness is large and independent of the antenna gap size (Supplementary Fig. 3C-D), therefore likely representing true nanoscale dynamic heterogeneity of receptors on the cell membrane.

To verify that our multiplexed approach is sensitive to differences in molecule dynamics in living cells, we measured fluorescent lipids and fluorescent ligand-labelled transmembrane proteins that diffuse faster and slower than FR-GPIs, respectively. Indeed, the median diffusion times of DPPE lipids recorded on the nanoantenna arrays (Fig. 3E) was faster than on FR-GPIs (Fig. 3F) while the seven-pass transmembrane protein ADORA2A, a GPCR labelled with a fluorescent ligand (NECA), displayed diffusion times that were markedly slower (Fig. 3G) than FR-GPIs. Altogether, our approach resolves different nanoscale dynamics of membrane lipids and proteins in living cells in a multiplexed, high throughput fashion.

To understand the absence of gap size-dependence on diffusion times and fluorophore brightness we performed FDTD simulations (Fig. 4A). The simulations predict that the lateral extent of the near-field emanating from the gap increases non-linearly as a function of the axial distance $z$ away from the antenna (Supplementary Fig. 4). At distances above 15 nm, the effective lateral confinement becomes comparable for all antennas regardless of their gap size (Supplementary Fig. 4). Moreover, the maximum field intensity relative to the aperture decays rapidly as a function of $z$, converging to similar values for $z > 10$ nm, for all gap sizes (Fig. 4B). As the relative brightness of single molecules measured from antennas is proportional to the maximum field intensity (which is valid for high quantum yield fluorophores as the ones used here), these predictions explain the lack of gap size-dependence on the living cell experiments (Supplementary Fig. 3C-D) and indicate that fluorophores in the cell membrane are located at axial distances above 10-15 nm from the nanoantennas.

Interestingly, these simulations imply that the fluorophore's axial positions could be accurately determined by exploiting the exquisite brightness-to-axial-distance relationship provided by antennas. For this, we used an empirical description to relate the axial distance from the antennas to the measured fluorophore brightness (Supplementary Note 2) and placed the experimental datasets of the beads and the FR-GPIs onto this description (Fig. 4C). Whereas beads explore the full 3D spatial confinement of the nanoantennas, FR-GPIs diffuse at ~20 nm above the antenna substrate, in agreement with their expected heights due to receptor lengths and the glycocalyx thickness[27]. Using this approach, we further extracted the axial distances from the plasma membrane of three sets of fluorescent probes. We found the lowest relative brightness for the Atto647N-conjugated lipid DPPE corresponding to a distance of 27±1 nm above the antenna (Fig. 4D), while for FR-GPIs and NECA, their average heights above the antenna substrates resulted 26±2 nm and 22±8 nm, respectively (Fig. 4D). Since the antennas are planar, our approach thus delivers accurate axial nanoscale information of molecular components within the plasma membrane of living cells.

Overall, our data demonstrate unique attributes by combining antenna arrays with multiplexed fluorescence readout: High-throughput data acquisition at the nanoscale, highly localised fluorescent enhancement and, simultaneous nanoscale dynamic readouts in biological relevant scenarios with high temporal resolution. Our method can be easily adapted onto existing epi-illumination setups, where interrogation volumes and number of antennas in the FOV are solely dependent on antenna design. Sensitivity and temporal resolution are currently limited by camera technology but will be a balance between the photon emission during each event and pixel noise at the detector. Numerous applications are envisaged; each nanoantenna provides a reaction volume to follow the progression of chemical reactions at the single-molecule level, allowing building up an ensemble of states from the multiplexed readout. The assessment of binding isotherms at micro-millimolar affinities may be determined, even in the presence of high concentrations of binding partners. An exciting application is its compatibility with live-cell experimentation on the cell membrane. We foresee that future fabrication optimisations such as tweaking the colour sensitivity and antenna-to-antenna spacing would allow for tailoring to a vast range of possibilities, in particular, to elucidate live-cell membrane dynamics at the nanoscale with high throughput.

## METHODS

### Planar gold nanoantenna array

Large-scale planar gold nanoantenna arrays with surface nanogaps were fabricated as reported earlier[14]. The antenna design was based on the "antenna-in-box" platform combining a central nanogap region between two 80 nm gold half-spheres, and a cladding 300×140 nm² box to screen the background by preventing direct excitation of molecules diffusing away from the nanoantenna gap. The individual nanoantennas were fabricated with gap sizes of 10, 20, 25, 30 and 35 nm and an antenna-to-antenna spacing of 4 μm. The multistep process to fabricate the nanoantenna arrays comprised electron-beam lithography (EBL), planarization, etch back, and template stripping. The planarization step rendered the full antenna substrate perfectly flat with height differences between the antenna regions and the surrounding gold substrate below 2 nm[14]. Before each experiment the antenna substrate was carefully cleaned by rinsing with ethanol and MilliQ water followed by $N_2$ drying. A final 3-5 min deep-UV plasma treatment rendered the gold surface hydrophilic promoting cell adhesion and spreading[28].

### Experimental setup

The fluorescence experiments were performed on a home-built setup built around a Nikon Eclipse Ti body. The laser beam from a 642 nm cw-laser (SpectraPhysics Excelsior-640C) was expanded and confined to provide a uniform beam waist of around 10 mm. After a half-waveplate for polarisation control, and via a set of lenses, the beam was directed into the back of the Nikon body and focused on the back focal plane of a CFI Apo TIRF 60×, 1.49 NA objective. The excitation was aligned in an epi-illumination configuration and the emission was collected via the same objective, separated from the excitation using a quad-edge dichroic (Semrock) in combination with a quad-notch filter (Semrock). The emission was further filtered through a 675/68 nm bandpass filter (Semrock) and finally directed onto a water-cooled Prime 95B sCMOS camera (Photometrics). The system was controlled using µManager software (v2.0 gamma)[29]. The planar nanogap antenna arrays were illuminated at 45-50 mW laser power at the back focal plane, corresponding to a power density at the sample plane of about 0.05 kW/cm$^2$. The wavelength of the excitation corresponds to the resonance of the nanogap antennas and the direction of the excitation polarisation (2500:1) was adjusted and aligned along the antenna dimer using a half-wave plate (Newport).

*Multiplexed burst analysis on nanoantennas*

To provide enough signal on the camera we first used bright far-red 45 nm size fluorescent beads (F8789, ThermoFisher) at a concentration of 24 nM and further slowed down their diffusion from 5.2±0.7 µm$^2$/s in a 20mM Tris buffer (pH 8) to 1.4±0.5 µm$^2$/s in a 1.6 M sucrose solution. A 5 ms integration time was used to establish a compromise between allowing enough time for an event to occur while preventing multiple events to be captured and integrated within a single frame. The 10 Hz framerate subsequently provided sufficient sampling for multiple events within a reasonable time. A total of 225 nanoantennas could be sampled simultaneously using a region of interest (ROI) of 570×570 pixels (122 nm pixel size). To remove the fixed pattern noise due to pixel-by-pixel variation in the amplification, the raw image from the camera was converted to photo-electrons using previously described methods[19]. The projected image from all the converted 2000 frames of the measurement showed that aside from the bright signal emerging from the diffraction-limited antennas there was an uneven background from the gold film reflection. Therefore, each raw image was background subtracted and flatfield corrected. The background was estimated per frame using a Gaussian filter of 30 pixels. The flatfield excitation correction was estimated from the projected image using the same filter setting. The (x,y) position of each antenna was identified and its intensity time trace was calculated by summing up the photo-electrons in a circular area with a radius of 4 pixels around the identified antenna position in each of the 2000 frames. A positive event of a high intensity burst arising from a diffusing bead was assigned when the measured number of photo-electrons of a trace were higher than the average plus 3 times the standard deviation. To reduce sticking effects of the beads on the nanoantennas that would bias the results, we used a high pH buffer rendering the bead and antenna surface negatively charged and we added Tween-20 (<1% of total volume, Sigma) to the buffer solution. Additionally, any high-intensity bursts occurring in adjacent frames, potentially resulting from sticking, were rejected for further analysis. Each trace typically displays more than 12 high intensity bursts (Supplementary Fig. 1).

*Multiplexed fluorescence correlation spectroscopy on nanoantennas*

Data for FCS was obtained similarly as described above, but adjusting the ROI and integration settings of the camera. With 50×950 pixels exactly 50 nanoantennas could be imaged simultaneously. At 0.9 ms integration time 24000 frames were collected at >1 kHz. All frames were converted to photo-electrons and antenna positions were identified from the projected image. The raw traces were autocorrelated using a multi-tau approach. Here the background signal and camera noise are not expected to contribute to the autocorrelation function (ACF) that is derived from the intensity trace. The autocorrelation offset however is lowered with increasing uncorrelated photons and the values obtained from the ACF fitting were corrected during post-processing. Other potential artefacts, such as photobleaching or sticking events, were removed by segmenting the trace into six equally long parts, from which six individual autocorrelation curves were calculated. The average ACF of the accepted traces for each antenna were further fitted using Quickfit3[30]. Given that the near-field of the nanoantenna is axially and laterally confining the detection volume, we found that the best and simplest diffusion model fitting the correlation curves is a 2D normal diffusion fit with a single component:

$$G(i, \tau) = \frac{1}{N_i} \cdot \frac{1}{(1 + \tau/\tau_{i,diff})}$$

with $N_i$ denoting the average number of molecules in the detection volume of antenna ($i$) and $\tau_{i,diff}$ the characteristic diffusion time measured at antenna ($i$). The overestimation of $N_i$ due to the high background, $\langle B_i \rangle$, was corrected using:

$$N_{i,corr} = N_i \cdot \left(\frac{\langle F_i \rangle - \langle B_i \rangle}{\langle F_i \rangle}\right)^2$$

Fluorophore brightness associated to antenna excitation was calculated by dividing the corrected average trace intensity ($\langle F_{i,corr} \rangle = \langle F_i \rangle - \langle B_i \rangle$) with the corrected number ($N_{i,corr}$): $Q_i = \frac{\langle F_{i,corr} \rangle}{N_{i,corr}}$.

The relative brightness was obtained through division by the median value acquired from the aperture recordings during the same measurement set.

*Cell culture and labelling*

The folate-receptor-expressing human gastric cell line AGS (FRAGS) was maintained in HF12 medium containing 50 mg/ml Hygromycin. Prior to the experiment cells were detached from the flask using TrypLE Express (ThermoFisher), collected and washed with M1 buffer (150 mM NaCl, 20 mM HEPES, 5 mM KCl, 1 mM $CaCl_2$ and 1 mM $MgCl_2$, pH = 7.4). Next, the cells were allowed to attach freely to a freshly cleaned antenna substrate in M1-Gl buffer (M1 buffer containing 2 mg/ml glucose) for 2 hours at 37 °C. The GPI-anchored folate receptors on the attached FRAGS cells were labelled with 200 nM Abberior Star635 conjugated Fab-fragments against the Folate receptor (Mov19, degree of labelling was 0.6). See Supplementary Note 3 for further details regarding the labelling of other membrane components investigated.

*Simulations*

Finite Different Time Domain (FDTD) Simulations were performed for planar gold dimer nanoantennas (dimer diameter 80 nm) for various gap sizes embedded in a aperture as approximately used in the experiments. Computations were made using the FDTD method in the RSoft Fullwave Software with a mesh size of 1 nm and 214 temporal steps of $8.1 \cdot 10^{-19}$ seconds[14]. The simulations were calculated at a laser excitation of 633 nm. From the generated simulation plots, we extracted quantitative information such as the maximum intensity and the corresponding lateral and axial extensions of the field.

**Acknowledgments:**

The authors are grateful to C. Prabhakara for generously providing the labelled Fab fragment together with the labelled nanobodies, and to both C. Prabhakara and R. Chandran for help with the Drosophila Hemocyte experiments.

**Author contributions:**

All the authors conceived the project. P.M.W. and T.S.v.Z. implemented the multiplexed optical scheme, performed experiments on the nanoantennas and analyzed the data. S.M. and M.F.G.-P supervised the research. All authors contributed to the interpretation of the results and the writing of the manuscript.

**Funding:**

The research leading to these results has received funding from the European Commission H2020 Program under grant agreement ERC Adv788546 (NANO-MEMEC) (to M.F.G.-P.), COFUND Doctoral Programme of the Marie- Skłodowska -Curie-Action of the European Union's Horizon 2020 research and innovation programme under the Marie Skłodowska-Curie grant agreement No 665884 (to P.M.W); Government of Spain (Severo Ochoa CEX2019-000910-S, State Research Agency (AEI) PID2020-113068RB-I00 / 10.13039/501100011033 (to M.F.G.-P.), Fundació CELLEX (Barcelona), Fundació Mir-Puig and the Generalitat de Catalunya through the CERCA program and AGAUR (Grant No. 2017SGR1000 to M.F.G.-P.). T.S.v.Z. acknowledges an EMBO fellowship (ALTF 1519-2013) and the NCBS Campus fellowship. S.M. acknowledges a JC Bose Fellowship from the Department of Science and Technology (Government of India), a collaborative grant from the Human Frontiers Science Program (RGP0027/2012 with M.F.G.-P.), and support from a Welcome Trust/Department of Biotechnology, Alliance Margdarshi Fellowship (IA/M/15/1/502018).

# REFERENCES

1. Moerner, W. E. New directions in single-molecule imaging and analysis. *Proceedings of the National Academy of Sciences* **104**, 12596–12602 (2007).
2. Doppagne, B. *et al.* Single-molecule tautomerization tracking through space- and time-resolved fluorescence spectroscopy. *Nat. Nanotechnol.* **15**, 207–211 (2020).
3. Weiss, L. E. *et al.* Three-dimensional localization microscopy in live flowing cells. *Nat. Nanotechnol.* **15**, 500–506 (2020).
4. Levene, M. J. *et al.* Zero-Mode Waveguides for Single-Molecule Analysis at High Concentrations. *Science* **299**, 682 (2003).
5. Altug, H., Oh, S.-H., Maier, S. A. & Homola, J. Advances and applications of nanophotonic biosensors. *Nat. Nanotechnol.* **17**, 5–16 (2022).
6. Kinkhabwala, A. *et al.* Large single-molecule fluorescence enhancements produced by a bowtie nanoantenna. *Nature Photonics* **3**, 654–657 (2009).
7. Novotny, L. & van Hulst, N. Antennas for light. *Nature Photonics* **5**, 83–90 (2011).
8. Punj, D. *et al.* A plasmonic 'antenna-in-box' platform for enhanced single-molecule analysis at micromolar concentrations. *Nature Nanotechnology* **8**, 512–516 (2013).
9. Acuna, G. P. *et al.* Fluorescence Enhancement at Docking Sites of DNA-Directed Self-Assembled Nanoantennas. *Science* **338**, 506–510 (2012).
10. Holzmeister, P., Acuna, G. P., Grohmann, D. & Tinnefeld, P. Breaking the concentration limit of optical single-molecule detection. *Chemical Society Reviews* **43**, 1014–1028 (2014).
11. Puchkova, A. *et al.* DNA Origami Nanoantennas with over 5000-fold Fluorescence Enhancement and Single-Molecule Detection at 25 μM. *Nano Letters* **15**, 8354–8359 (2015).
12. Goldschen-Ohm, M. P., White, D. S., Klenchin, V. A., Chanda, B. & Goldsmith, R. H. Observing Single-Molecule Dynamics at Millimolar Concentrations. *Angew. Chem. Int. Ed.* **56**, 2399–2402 (2017).
13. Trofymchuk, K. *et al.* Addressable nanoantennas with cleared hotspots for single-molecule detection on a portable smartphone microscope. *Nat Commun* **12**, 950 (2021).
14. Flauraud, V. *et al.* In-Plane Plasmonic Antenna Arrays with Surface Nanogaps for Giant Fluorescence Enhancement. *Nano Letters* **17**, 1703–1710 (2017).
15. Flauraud, V. *et al.* Large-Scale Arrays of Bowtie Nanoaperture Antennas for Nanoscale Dynamics in Living Cell Membranes. *Nano Letters* **15**, 4176–4182 (2015).
16. Lohmüller, T. *et al.* Single Molecule Tracking on Supported Membranes with Arrays of Optical Nanoantennas. *Nano Letters* **12**, 1717–1721 (2012).
17. Winkler, P. M. *et al.* Transient Nanoscopic Phase Separation in Biological Lipid Membranes Resolved by Planar Plasmonic Antennas. *ACS Nano* **11**, 7241–7250 (2017).
18. Regmi, R. *et al.* Planar Optical Nanoantennas Resolve Cholesterol-Dependent Nanoscale Heterogeneities in the Plasma Membrane of Living Cells. *Nano Letters* **17**, 6295–6302 (2017).
19. Huang, F. *et al.* Video-rate nanoscopy using sCMOS camera–specific single-molecule localization algorithms. *Nature Methods* **10**, 653–658 (2013).
20. Yuan, H., Khatua, S., Zijlstra, P., Yorulmaz, M. & Orrit, M. Thousand-fold Enhancement of Single-Molecule Fluorescence Near a Single Gold Nanorod. *Angewandte Chemie International Edition* **52**, 1217–1221 (2013).
21. Kim, S. A., Heinze, K. G. & Schwille, P. Fluorescence correlation spectroscopy in living cells. *Nat Methods* **4**, 963–973 (2007).

22. Ohlídal, I. & Lukeš, F. Optical analysis of thin gold films by combined reflection and transmission ellipsometry. *Thin Solid Films* **85**, 181–190 (1981).
23. Honigmann, A. *et al.* Scanning STED-FCS reveals spatiotemporal heterogeneity of lipid interaction in the plasma membrane of living cells. *Nature Communications* **5**, 5412 (2014).
24. Goswami, D. *et al.* Nanoclusters of GPI-Anchored Proteins Are Formed by Cortical Actin-Driven Activity. *Cell* **135**, 1085–1097 (2008).
25. Nohe, A., Keating, E., Fivaz, M., van der Goot, F. G. & Petersen, N. O. Dynamics of GPI-anchored proteins on the surface of living cells. *Nanomedicine: Nanotechnology, Biology and Medicine* **2**, 1–7 (2006).
26. Gowrishankar, K. *et al.* Active Remodeling of Cortical Actin Regulates Spatiotemporal Organization of Cell Surface Molecules. *Cell* **149**, 1353–1367 (2012).
27. Kanchanawong, P. *et al.* Nanoscale architecture of integrin-based cell adhesions. *Nature* **468**, 580–584 (2010).
28. Cai, S. *et al.* Recent advance in surface modification for regulating cell adhesion and behaviors. *Nanotechnology Reviews* **9**, 971–989 (2020).
29. Edelstein, A., Amodaj, N., Hoover, K., Vale, R. & Stuurman, N. Computer Control of Microscopes Using μManager. *Current Protocols in Molecular Biology* **92**, (2010).
30. Krieger, J. W. & Langowski, J. QuickFit 3.0. *http://www.dkfz.de/Macromol/quickfit/*.

**FIGURE CAPTIONS**

**Figure 1: Multiplexed nanogap antenna detection. (A)** Schematics of the experimental configuration. Nanogap antenna substrates are excited in an epi-configuration. Emission from freely diffusing fluorescent beads within the antenna gaps is collected with an sCMOS camera. The detected signal increases in the event of a bead diffusing through the antenna during a 5 ms frame (bottom left sketch). A time-averaged 2000 frame movie displays a total of 225 antennas addressed simultaneously. Antennas are visible as diffraction-limited spots. **(B)** Representative fluorescent intensity bursts (red) and corresponding intensity histograms (right) from a 2000 frame time series (background corrected) of fluorescent beads (d ~ 45 nm, 24 nM) in a 1.6 M sucrose solution obtained on a single antenna of 25 nm gap size, excited with a polarisation perpendicular (black, out of resonance) or parallel to the gold dimer (red, in resonance). **(C)** Normalised ensemble photon count distribution of 225 nanogap antennas measured simultaneously for different gap sizes, together with the apertures as a control. **(D)** Peak brightness (colour-encoded) for different antenna arrays, each containing 225 antennas of similar gap size. Each pixel represents a single nanogap antenna. Each set displays nanogap antennas of decreasing gap sizes. A set of nine aperture arrays (each containing 25 apertures) is also included as a control for the maximum peak brightness in absence of antenna near-field enhancement. **(E)** Relative fluorescent enhancement for the top 1% performing nanogap antennas. The exponential decay as the gap sizes increase is for visual guidance. Each dot corresponds to an individual antenna.

**Figure 2. Simultaneous multiplexed nanoantenna fluorescence correlation spectroscopy. (A)** A time-averaged field-of-view of 50 nanogap antennas shown as bright spots, imaged at 1 kHz camera framerate. **(B)** Bead brightness as function of the radius used for photon collection around the antenna, from 50 nanoantennas of 25 nm gap size. **(C)** Autocorrelation functions (ACFs) of diffusing beads recorded on a single 25 nm nanogap antenna (red) at 1 kHz framerate, and signal collected from a region (of equal size) on the gold film (grey). The average (dots) and standard-deviation (shaded area) of six sub-traces from a single trace (24 s) are shown. **(D)** ACF-carpet of diffusing beads on nanoantennas of different gap sizes. The carpet displays 184 ACFs from multiple measurements, arranged vertically and grouped according to antenna gap size, as indicated on the left. Each row in the ACF-carpet represents the colour-encoded time evolution of the normalised ACF amplitude from a single antenna. On the left of the full ACF-carpet, the first three lag times of the raw ACFs are displayed and colour-coded according to the raw correlation amplitude G(t). **(E)** Relative bead brightness of the fluorescence signals from antennas of different gap sizes as compared to the signal from the aperture, obtained from fitting the raw ACFs curves displayed in (D). The relative brightness distribution from aperture recordings is included for comparison. Each dot represents a single antenna or aperture. **(F)** Diffusion times obtained from fitting the raw ACFs curves displayed in (D). Each dot represents a single antenna.

**Figure 3. Multiplexed FCS at the nanoscale of fluorescently labelled biomolecules in the plasma membrane of living cells. (A)** A camera snapshot of cells (outlined in white) spread onto the nanoantenna array platform, with the nanoantennas visible as diffraction-limited bright spots. Scale bar is 10 μm. **(B)** ACF-carpet of FR-GPIs diffusing on living cells for different nanoantennas. Each row represents one ACF from an individual antenna where the correlation amplitude G(t) is colour-encoded and the antenna gap size is indicated on the left. The carpet displays a total of 51 autocorrelation curves from 2 independent measurements. **(C)** Examples of one-component fittings of raw ACFs from FR-GPI fluorescence intensity traces over three different 25 nm gap antennas. **(D)** Scatter plot of diffusion time versus relative brightness as extracted from ACF curves of FR-GPIs diffusing above nanoantennas of different gap sizes (purple) or apertures (grey). Each dot corresponds to an individual antenna or aperture measurement. **(E-G)** Diffusion time distributions of Atto647N conjugated DDPE lipids (E), Abberior star635-Fab labeled FR-GPI (F) and NECA-Bodipy630 labeled GPCR (ADORA2A) (G). Each dot corresponds to an individual antenna measurement. The median diffusion times are: 6.5 ms (E), 11.7 ms (F) and 39.5 ms (G).

**Figure 4. FDTD simulations and empirical description of the axial near-field of planar antennas for different gap sizes. (A)** Near-field intensity profiles along z-x cuts (left) and z-y cuts (right) for antenna gap sizes of 15 (top row) and 25 nm (bottom row). The same intensity scale is used for all the plots to allow for comparison. **(B)** Normalised maximum field intensity relative to the aperture of the z-x (dot symbols) and z-y (triangle symbols) cuts as a function of the axial direction, i.e., z heights above the x-y plane of the nanoantenna (z = 0 nm), for gap sizes indicated in the legends. The median normalised field intensity of all the antenna gap sizes is shown in magenta. **(C)** Fitted median of the normalised maximum field intensity from the different gap sizes as calculated in (B) and extrapolated until z = 30 nm away from the antenna plane (z = 0 nm). The empirical relationship was used to estimate the height above the antenna plane using the brightness values of the experimental data obtained on fluorescent beads and FR-GPI in living cells on nanoantennas. **(D)** Relative brightness (left) and calculated height above the antenna substrate (right) of DPPE, FR-GPI and NECA-labelled GPCR. The distribution of relative brightness for the aperture has been included for comparison. Each dot corresponds to an individual antenna or aperture measurement.

FIGURE 1

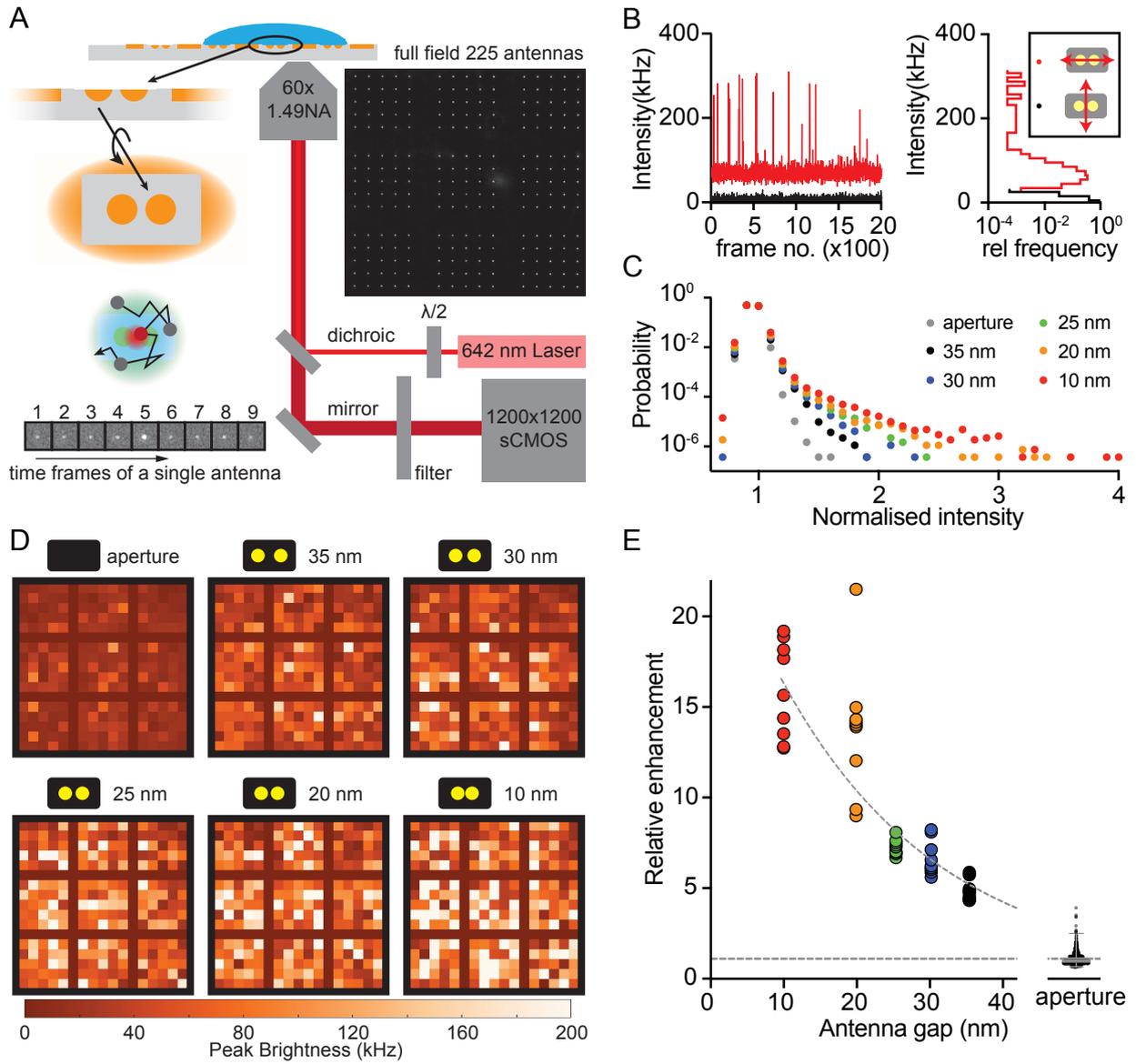

FIGURE 2

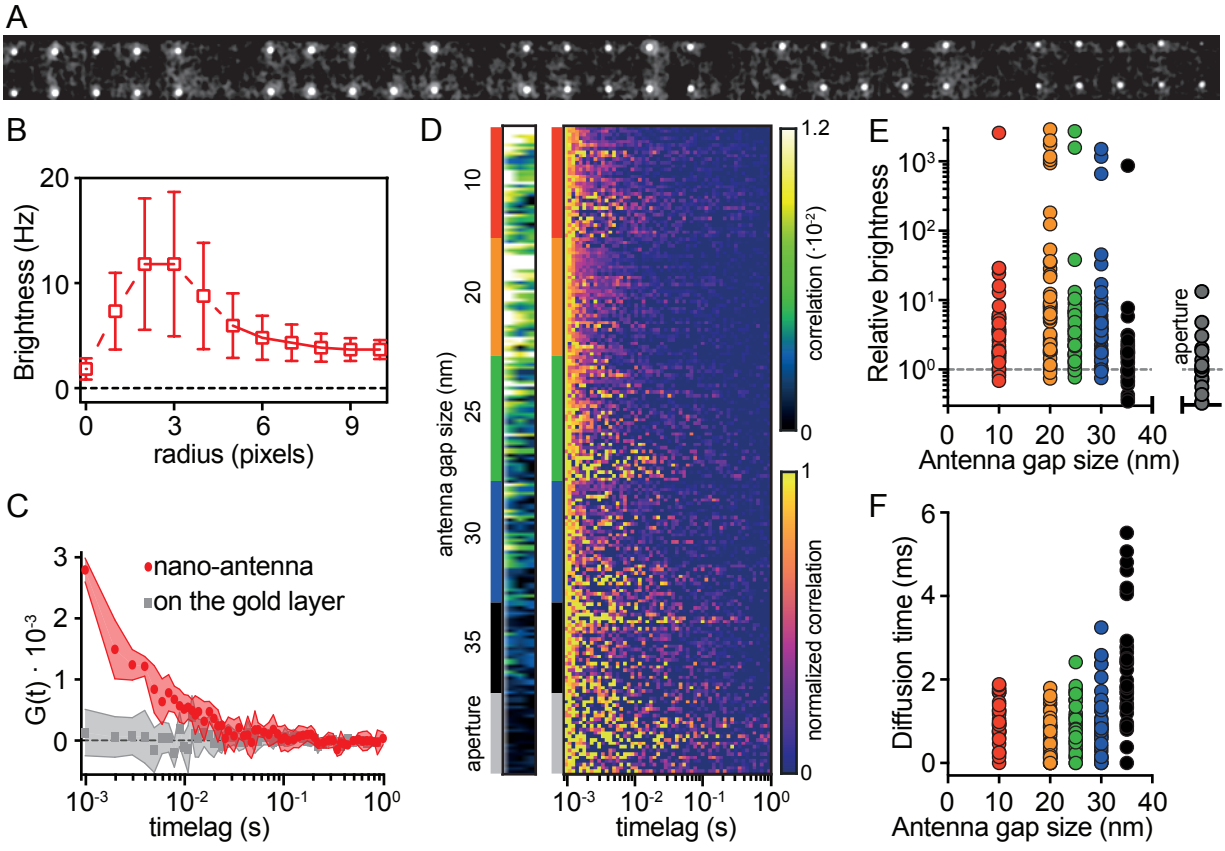

FIGURE 3

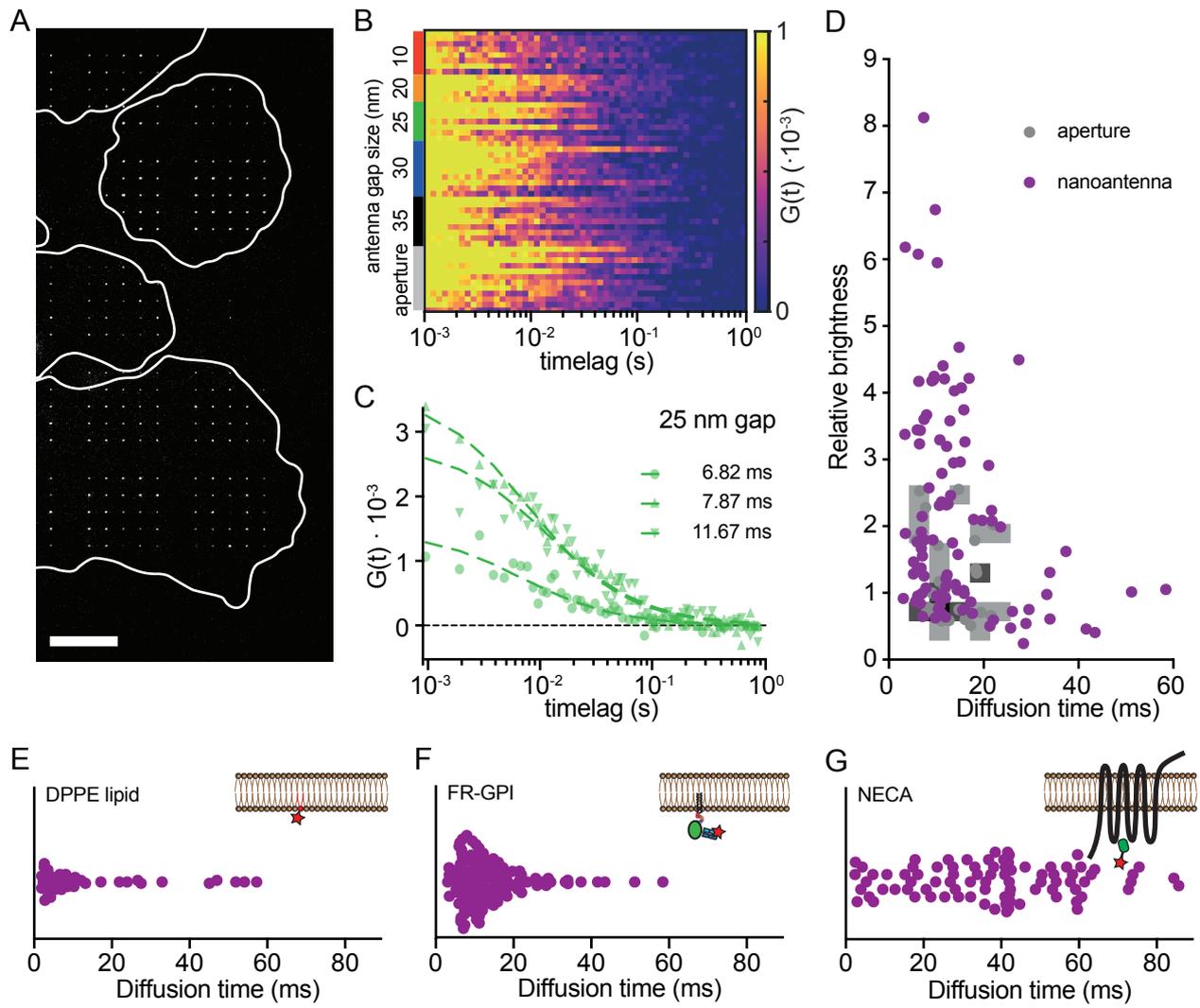

FIGURE 4

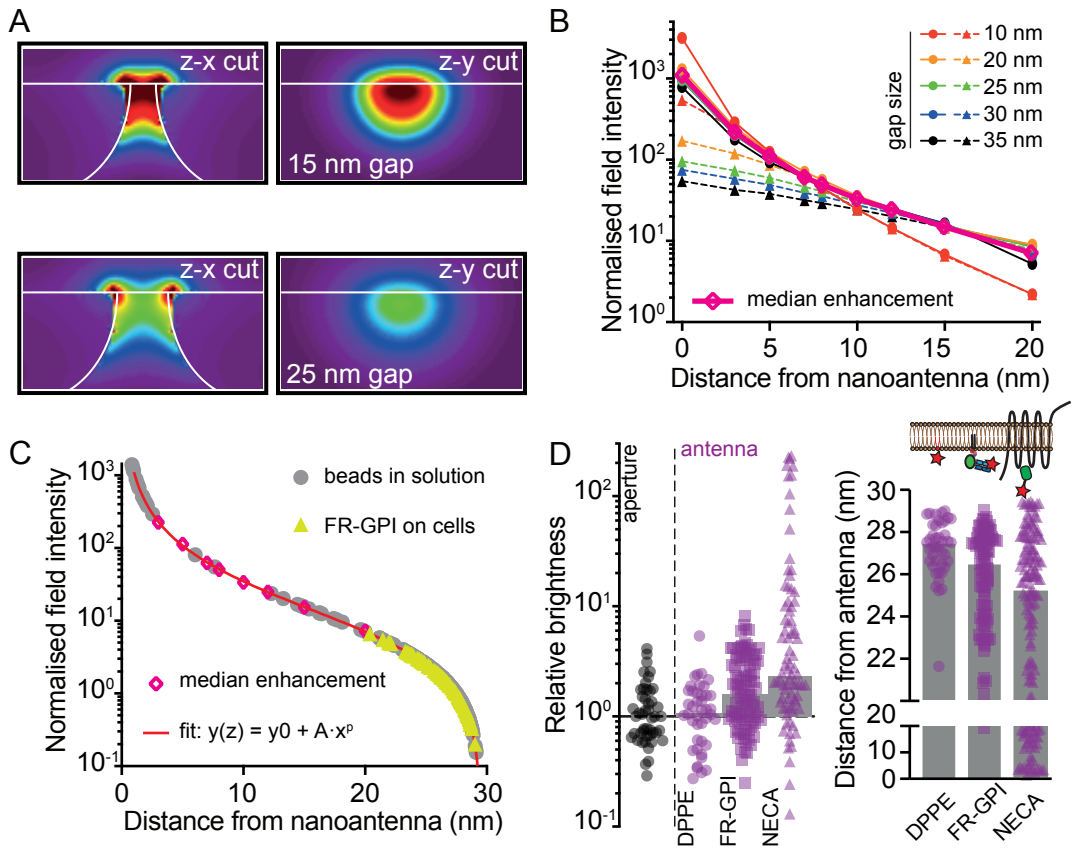